\documentstyle[preprint,tighten,prb,aps]{revtex}     %

\begin{document}
\draft

\preprint{RCTP-9402, gr-qc/9408029}

\author{J.E. Rankin\thanks{%
E-mail address: (Internet) jrankin@panix.com}}
\address{Rankin Consulting, 527 Third Avenue, Ste. 298,
New York, NY 10016, USA}

\title{QUANTUM BEHAVIOR IN ASYMMETRIC, WEYL-LIKE CARTAN GEOMETRIES
\footnote{Originally given as a talk at the MG7 Meeting, 24-29 July, 1994.}}

\date{August 31, 1994}

\maketitle

\begin{abstract}
This discussion examines recent developments in the theory of a Weyl-like,
Cartan geometry with natural Schr\"odinger field behavior proposed
previously. In that model, very nearly exactly a coupled Einstein-Maxwell-
Schr\"odinger, classical field theory emerges from a gauge invariant, purely
geometric action based solely on variations of the electromagnetic potentials
and the metric. In spite of this, only slight differences appear between the
resulting Schr\"odinger part, and the conventional theory of the
Schr\"odinger field. Close examination of the differences reveals that most
are general relativistic effects which are unobservable in flat spacetime,
and which are estimated to interact significantly only via their
gravitational fields, or on scales comparable with neutrino interaction cross
sections. The only remaining difference is that the wavefunction obeying the
conjugate wave equation is not always restricted to be exactly the complex
conjugate of the primary wavefunction. Generalizations of the model lead
naturally to spinlike phenomena, a possible new mechanism for a theory of
rest mass, and spinor connections containing the form of an SU(2) potential.

\end{abstract}

\section{Introduction}

In 1927, London published his famous paper on the quantum mechanical
interpretation of Weyl's classical geometry\cite{london.orig}. In the last 15
years, a flurry of new articles has appeared, reviving interest in intrinsic
quantum properties of both Weyl geometries, and closely related geometric
structures\cite{rankin.ijtp,galehouse.ijtp,wheeler.physrev,rankin.caqg,%
wood.papini,rankin.preprint}. In this particular discussion, I will elaborate
on a few points mentioned in a recent paper on such a case, a purely field
theoretic model cast in a Weyl-like Cartan geometry with {\it intrinsic}
Schr\"odinger field behavior\cite{rankin.preprint}. Indeed, this discussion
should be viewed as an additional section of comments in that paper.

One reason this model is of interest is because it yields essentially a full,
coupled Einstein-Maxwell-Schr\"odinger set of fields from a purely geometric
action in which only the Weyl four vector and the metric are varied. Yet the
Schr\"odinger field also emerges from the results, correctly obeying the
Klein-Gordan equation and a charge conjugate, complementary equation, and
producing the correct Schr\"odinger current and stress tensor as sources of
the electromagnetic and gravitational fields. Additional terms do appear in
the equations, but they will be noted to produce vanishing effects in the
limit of a flat spacetime of special relativity. Thus, they are essentially
general relativistic modifications to special relativistic Schr\"odinger
theory. Addition of further geometric fields to the model produces spinlike
phenomena, and terms which have the form of an SU(2) potential.

\section{General Relativistic Nature of Additional Terms}

In the earlier reference ($\hat \phi =\pi $ case)\cite{rankin.preprint}, the
Schr\"odinger wave function $\psi $ obeys the wave equation
{\samepage
\begin{eqnarray}
& &(1/\sqrt {-\hat g}\, )(\sqrt {-\hat g}\, \hat g^{\mu \nu }
\psi _{,\nu })_{,\mu }+2\hat g^{\mu \nu }v_\mu \psi _{,\nu }
\nonumber \\
& &+(1/\sqrt {-\hat g}\, )(\sqrt {-\hat g}\, \hat g^{\mu \nu }
v_\nu )_{,\mu }\psi +\hat g^{\mu \nu }v_\mu v_\nu \psi =
-{\textstyle{1 \over 6}}\, (1+\hat R)\psi
\label{sch.wave.eq}
\end{eqnarray}}%
where $v_\mu $ is the Weyl vector, and is proportional to the standard
Maxwell potential through an imaginary coefficient. A conjugate wave
equation (minus $v_\mu $) governs the conjugate wavefunction $\xi $. The
formalism does {\it not} require $\xi $ always to be the complex conjugate of
$\psi $. Except for that, the Schr\"odinger current and stress tensor contain
the standard terms\cite{morse.feshbach}. However, the stress tensor also
contains additional terms $\hat \beta _{||\gamma }^{\: \: \: \: \; ||
\gamma }\hat g_{\mu \nu }-\hat \beta _{||\mu ||\nu }$, where $\hat \beta =
\xi \psi $, and the $||$ derivative is the covariant derivative using the
Christoffel symbols formed from $\hat g_{\mu \nu }$. These terms are in
addition to {\it manifestly} general relativistic modifications, such as the
$\hat R $ term in Eq. (\ref{sch.wave.eq}).

However, terms in the stress tensor must interact with their environment in
some way in order to be observable. In that regard, even though
$\hat \beta _{||\gamma }^{\: \: \: \: \; ||\gamma }\hat
g_{\mu \nu }-\hat \beta _{||\mu ||\nu }$ appear capable of carrying energy
and momentum magnitudes that might be comparable to the conventional
Schr\"odinger terms, these terms can only be detected insofar as their
divergence is nonvanishing (allowing energy exchange with other terms), or
via the gravitational field they produce. But the gravitational field is a
manifestly general relativistic effect. And, the divergence of the terms is
\begin{equation}
[\hat \beta _{||\gamma }^{\: \: \: \: \; ||\gamma }\delta _\mu ^\nu -\hat
\beta _{||\mu }^{\: \: \: \: \; ||\nu }\, ]_{||\nu }=
\hat R_\mu ^{\: \: \nu}\hat \beta _{,\nu }
\label{div.extra}
\end{equation}
Thus, the interaction of these terms with the rest of the stress tensor is
{\it also} a manifestly general relativistic effect (more precisely, it is a
higher order effect). The terms are therefore completely unobservable in the
limit of the flat spacetime of special relativity.

But given that these terms are therefore general relativistic, when might
they, or other general relativistic effects be important? Of course, they
might contribute significantly to overall gravitational fields in any almost
any configuration. But beyond that, examination of Eq. (\ref{div.extra})
suggests that whenever the magnitude of $\hat R_{\mu \nu }$ approaches or
exceeds unity, these modifications to standard Schr\"odinger theory may
become as significant as the usual kinetic energy and rest mass terms in the
stress tensor.

To obtain some crude estimate of when such cases arise, assume the Reissner-
Nordstr\"om metric\cite{adler} gives at least reasonable magnitude estimates
of $\hat R_{\mu \nu }$ in the vicinity of electrons, quarks, and other nearly
pointlike, fundamental particles. Further assume that the natural charge
magnitude associated with the problem is the electronic charge, and that the
natural length scale (all quantities in these equations are kept
dimensionless) is $\hbar /(\sqrt {6}\, m_0 c)$, where $m_0 $ is the electron
rest mass. This appears to be the natural rest mass to use here rather than
the Planck mass because later extensions of this structure make it easy to
generate masses {\it larger} than some reference mass, but hard to generate
masses smaller than that mass\cite{rankin.preprint}. Thus, the natural
reference mass to introduce is one of the very smallest, nonzero masses
known, in this case, the electron rest mass.

Using the Reissner-Nordstr\"om solution then as a very rough guide, one finds
that to lowest order,
$\hat R_{\mu \nu }\sim 12[(Gm_0 ^2 )/(\hbar c)][e^2 /(\hbar c)](1/r^4 )$, and
that this approaches unity for a radius $r_0 =5.6\cdot 10^{-23}\; cm.$. While
this is noticeably smaller than the current resolution of the pointlike
nature of electrons or quarks\cite{texas92}, which is about
$10^{-17}\; cm.$, it is still huge compared to the Planck scale. Indeed, if
$r_0$ is translated into a very crude estimate of a cross
section\cite{schiff} by forming
$\pi r_0 ^2$, this is found to be $\sim 10^{-44}\; cm.^2 $, a value actually
comparable to the low end of observed neutrino cross sections\cite{texas92}.
And, some additional general relativistic terms in the equations will also
begin to be important under the same conditions.

\section{Comments on an Extended Formalism}

The previous paper also proposes enhancements to the proposed model via
inclusion of a self dual antisymmetric part to the metric,
$\hat a_{\mu \nu }$ (for spinlike phenomena), and a tracefree part to the
torsion\cite{rankin.preprint}. For such cases, the results suggest that the
effective rest mass in the wave equations is determined via
$m^2 _0 \sim -e^{-i\hat \phi }(1+{\textstyle{1 \over 4}}\, %
\hat a)$, where $\hat \phi $ is a constant angle, and
$\hat a=\hat a^{\mu \nu }\hat a_{\mu \nu }$. This form illustrates why it
would be difficult to produce rest masses smaller than some reference mass,
since that would require $\hat a\approx -4$. On the other hand, $\hat a$
tends to be complex, which should mimic the effects of a complex scattering
potential\cite{schiff}, creating wavefunction sources and sinks (but without
violating conservation laws here). It remains to be seen if this can be
controlled and used to advantage to require something like a real eigenvalue
spectrum for $\hat a$ by forbidding pathological behavior. If not, it may be
necessary to simply constrain $\hat a$ to be real.

But when $\hat a_{\mu \nu }\neq 0$, previous results also suggest that it
generates the tracefree part to the torsion (the vector $\hat v_\mu $ is
$2/3$ the trace of the torsion). Indeed, if the tracefree part of the torsion
is constrained to be totally antisymmetric, the forms show a similarity to a
theory of torsion and spin developed by Hammond\cite{hammond}. Unfortunately,
this constrained case is omitted in deference to the more general case in the
earlier reference\cite{rankin.preprint}.

Another point associated with a tracefree part to the torsion
(denoted by $\hat Q^{\; \; \; \; \: \mu }_{\nu \alpha }$) concerns the form
of the spinor connection associated with the affine connection of the four
space. In this case, the affine connection is presumed to have a complex,
non-Christoffel (tensor) portion, so the spinor connection can be formed from
either the regular four space connection, or from its complex conjugate. In
practice, it appears necessary to mix both forms into some expressions in
order to have consistent results. However, rather than elaborate on that
here, simply assume that the spinor connection is defined via the usual
relation\cite{carmeli}
\begin{equation}
\hat \Gamma ^A _{B \alpha }=-{\textstyle{1 \over 2}}\, \hat g_{\mu \nu }
\sigma ^\nu _{E^\prime B}[\sigma ^{\mu E^\prime A}_{\; \; \; \; \; \; \;
\; \; ,\alpha }+\sigma ^{\gamma E^\prime A}\hat
\Gamma ^\mu _{\gamma \alpha }]
\label{def.spinor.conn}
\end{equation}

While space will not permit much detail here, the above expression can be
reworked by reexpressing
$2\sigma ^\nu _{E^\prime B}\sigma ^{\gamma E^\prime A}$ as the sum of two
terms, $\hat g^{\nu \gamma }\delta ^A _B $ symmetric in $\nu $ and $\gamma $,
and $S^{\nu \gamma \; \: A}_{\; \; \; \: B}$ antisymmetric (and self dual) in
the same index pair. Utilizing properties of self dual
expressions\cite{rankin.caqg}, Eq. (\ref{def.spinor.conn}) becomes
\begin{equation}
\hat \Gamma ^A _{B \alpha }=\{ ^{\: \hat A }_{B \alpha } \}-
{\textstyle{1 \over 2}}\,
S^{\; \; \gamma \; \: A}_{\alpha \; \: B}(\hat v_\gamma -i3\hat W_\gamma )
+{\textstyle{1 \over 2}}\, S^{\mu \gamma \; \: A}_{\; \; \; \: B}
\hat Q_{\mu \gamma \alpha }
\label{reordered.conn}
\end{equation}
where $\hat W_\gamma $ is the dual of the totally antisymmetric part of
$\hat Q_{\mu \gamma \alpha }$. Thus, a portion of the tracefree part of the
torsion is seen to mix directly with the four vector, $\hat v_\mu $.

Furthermore, since the term $S^{\mu \gamma \; \: A}_{\; \; \; \: B}$ is self
dual in the tensor indices, the term
${\textstyle{1 \over 2}}\, S^{\mu \gamma \; \: A}_{\; \; \; \: B}
\hat Q_{\mu \gamma \alpha }$ can be rewritten as
${\textstyle{1 \over 4}}\, S^{\mu \gamma \; \: A}_{\; \; \; \: B}
(\hat Q_{\mu \gamma \alpha }+i\: {}^* \hat Q_{\mu \gamma \alpha })$ where the
dual is taken on the first two indices of
$\hat Q_{\mu \gamma \alpha }$. But this is now the inner product of two self
dual forms, so that it can be expressed as the three space dot product of two
complex three vectors (with additional free spinor and tensor indices in this
case)\cite{rankin.caqg}. But the three vector corresponding to
$S^{\mu \gamma \; \: A}_{\; \; \; \: B}$ is just the Pauli matrices. The
overall result then is that
\begin{equation}
\hat \Gamma ^A _{B \alpha }=\{ ^{\: \hat A }_{B \alpha } \}-
{\textstyle{1 \over 2}}\,
S^{\; \; \gamma \; \: A}_{\alpha \; \: B}(\hat v_\gamma -i3\hat W_\gamma )
+\sum_{n} \tau_n b_{n \alpha }
\label{su2.form}
\end{equation}
where the $\tau_n $ are the Pauli matrices. But the last term now is in the
usual form for an SU(2) potential\cite{carmeli}. The fact that the preceding
term is {\it not} in the standard form for the U(1) potential in electroweak
theory does not appear so serious, since the purely U(1) case here is already
the cleanest form of this model.

\acknowledgments

I would like to thank Jim Nester and Jim Wheeler for helpful discussions.

\end{document}